\documentclass[AMA,STIX2COL]{WileyNJD-v2}

\articletype{Research in brief}%

\received{XX.XX.XXXX}
\revised{YY.YY.YYYY}
\accepted{ZZ.ZZ.ZZZZ}

\begin{document}

\title{Double arcsine transform not appropriate for meta-analysis}

\author[1]{Christian R\"{o}ver}
\author[1]{Tim Friede}

\authormark{C.~R\"{O}VER, T.~FRIEDE}

\address[1]{\orgdiv{Department of Medical Statistics}, \orgname{University Medical Center G\"{o}ttingen}, \orgaddress{\state{G\"{o}ttingen}, \country{Germany}}}

\corres{Christian R\"{o}ver, \email{christian.roever@med.uni-goettingen.de}}

\abstract[Abstract]{The variance-stabilizing Freeman-Tukey double arcsine transform was originally proposed for inference on single proportions. Subsequently, its use has been suggested in the context of meta-analysis of proportions. 
While some erratic behaviour has been observed previously, here we point out and illustrate general issues of monotonicity and invertibility that make this transform unsuitable for meta-analysis purposes.}


\jnlcitation{\cname{%
\author{C.~R\"{o}ver},
\author{T.~Friede}}
(\cyear{2022}), 
\ctitle{Double arcsine transform not appropriate for meta-analysis}, \cjournal{\href{https://arxiv.org/abs/2203.04773}{arXiv:2203.04773}}, \cvol{2022}.}

\maketitle


\section{Introduction}
  Freeman and Tukey (1950)\citep{FreemanTukey1950} proposed the \emph{double arcsine transform} in order to facilitate confidence interval construction or testing of \emph{single} proportions based on a normal approximation on the transformed scale.  Within a binomial model, for a given number~$a$ of events among a total of $N$~trials (i.e., an observed \emph{proportion} of $p=\frac{a}{N}$), the transformation is given by
  \begin{equation}
    \theta \; = \; \frac{1}{2}\biggl(\operatorname{arcsin}\Bigl(\sqrt{\textstyle\frac{a}{N+1}}\,\Bigr) + \operatorname{arcsin}\Bigl(\sqrt{\textstyle\frac{a+1}{N+1}}\,\Bigr)\biggr) \mbox{.}
  \end{equation}
  The double arcsine transform is a \emph{variance stabilizing transformation}, i.e., the variance of transformed deviates is (at least approximately) independent of the underlying binomial probability.\citep{Guan2009} Differing conventions are common regarding the overall scaling; above, a factor~$\frac{1}{2}$ was applied, but other definitions are also used.\citep{FreemanTukey1950,Guan2009,SchwarzerEtAl2019}
  It is worth also pointing out the closely related \emph{arcsine transform} here, which is defined as
  \begin{equation}
    \vartheta \; = \; \operatorname{arcsin}\Bigl(\sqrt{\textstyle\frac{a}{N}}\,\Bigr)
              \; = \; \operatorname{arcsin}\bigl(\sqrt{p}\,\bigr)\mbox{,}
  \end{equation}
  and which constitutes the limiting case for large sample sizes~$N$.\citep{Anscombe1948}
  While the (single) arcsine transform had been introduced in the 1930s,\citep{Bartlett1936,EisenhartCh16} Freeman and Tukey proposed the double arcsine variant suggesting improved variance stabilizing properties;\citep{FreemanTukey1949,FreemanTukey1950,Blom1954} The transforms' differing variances are also illustrated in this note's supplementary material.
  
  In the context of meta-analysis, Miller (1978)\citep{Miller1978} suggested the use of the double arcsine transform, and also worked out the inverse of the transform in order to back-transform combined estimates to the probability scale, which is essential for interpretation of the results. While the forward transformation is unambiguous, back-transformation requires specification of a corresponding ``sample size''~$N$\@. Miller (1978) suggested the harmonic mean of the original sample sizes here (without providing a rationale for this choice), and alternative conventions have been 
  proposed in the meantime.\citep{LinXu2020}
  Some general properties of arcsine transformation methods and related alternatives have been discussed by Lin and Xu (2020).\citep{LinXu2020}

  Schwarzer \emph{et~al.} (2019)\citep{SchwarzerEtAl2019} have previously pointed out certain pathologies arising when performing meta-analyses of prevalences on the double arcsine transformed scale and subsequently back-transforming to the probability scale. In particular, they demonstrated that one may end up with a zero estimate along with a zero-width confidence interval despite having observed positive proportions in the original data.
  In the present note, we will more closely investigate the double arcsine transform's properties and the origins of pathologies arising from its use in the meta-analysis context.


\section{Properties of the double arcsine transform}\label{sec:issues}
\begin{figure}
    \centering
    \makebox{\includegraphics[width=0.90\linewidth]{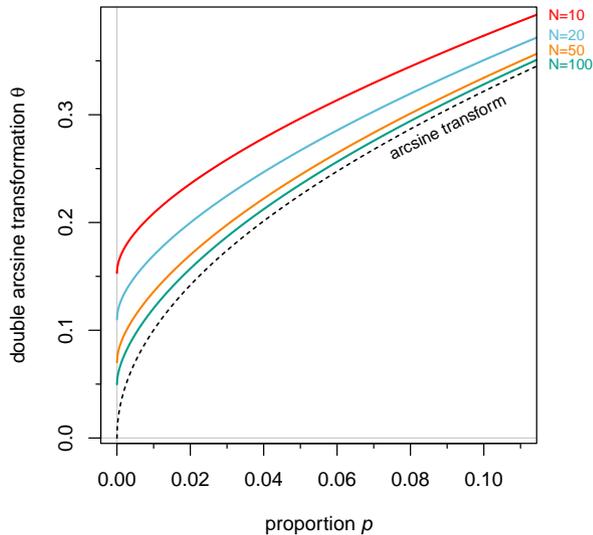}}
    \caption{\label{fig:trafo}Illustration of the double arcsine
      transform for varying sample sizes~$N$. Effectively,
      proportions (or proportion \emph{estimates}) corresponding to
      different sample sizes are not mapped to a common scale; even
      the ranges of possible $\theta$~values are only partly overlapping. The dashed line shows
      the limiting (``single'') arcsine transform.}
  \end{figure}

  In Figure~\ref{fig:trafo}, the double arcsine transform is sketched for selected values of~$N$, and in addition the limiting case (for large~$N$) of the arcsine transform is also shown.  
  \begin{table*}[b]
    \caption{\label{tab:example} Paradoxical behaviour of the
      double arcsine transform exemplified in the example data
      reported by Schwarzer \emph{et~al.}  (2019);\citep{SchwarzerEtAl2019}
      the transformed values~($\theta$) are not proportional to the
      original proportions~($p$) here (see also Figure~\ref{fig:example}).}  \centering
    \begin{tabular}{lcccc}
      \toprule
               & cases~($a$) & total~($N$) & proportion~($p$) & transformation~($\theta$) \\ 
      \midrule
      study~10 & $32$ & $16\,557$ & $0.00193$ & $0.0443$ \\
      study~13 & $\phantom{0}1$ & $\phantom{00}\,676$ & $0.00148$ & $0.0464$ \\
      \bottomrule
    \end{tabular}
  \end{table*}
  For \emph{fixed}~$N$, the transform is bijective and monotonic, but once differing $N$~values are considered, it becomes clear that proportions~$p$ are effectively mapped to differing scales.  Even the resulting \emph{images}, i.e., the ranges of $\theta$~values that the proportions are mapped to, are only partly overlapping.
  These issues do not arise for the limiting (``single'') arcsine transform.
  
  When considering differing sample sizes simultaneously, one of the consequences is that the transformation from~$p$ to~$\theta$ in general is not monotonic; an example is given in Table~\ref{tab:example}. These are results of two studies (``study~10'' and ``study~13'') from the example data quoted by Schwarzer \emph{et~al.} (2019).\citep{SchwarzerEtAl2019} While \emph{study~10} observed a greater proportion~$p$ of events than \emph{study~13}, the order is reversed on the transformed scale ($\theta$). 
  \begin{figure}[b]
    \centering
    \makebox{\includegraphics[width=0.90\linewidth]{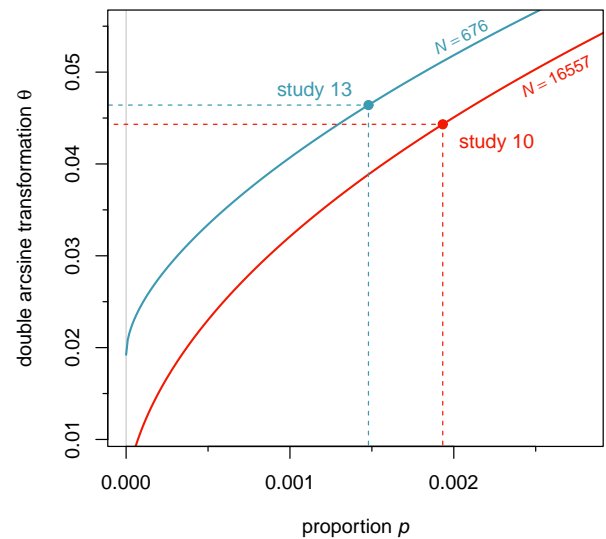}}
    \caption{\label{fig:example}Illustration of the double arcsine
      transformation for the example data from Table~\ref{tab:example}.}
  \end{figure}

  Figure~\ref{fig:example} illustrates the two differing transformations that are effectively applied for the two proportions, resulting in the reverse order on the transformed scale.  
  In such a setup, the combined estimate may eventually also not end up \emph{between} the observed proportions after back-transformation; for the example data from Table~\ref{tab:example}, this happens e.g. when using the inverse-variance convention\citep{BarendregtEtAl2013} to determine the $N$~value used for back-transformation (here: $N=17234$), resulting in an estimated proportion of $\hat{p}=0.00194$, which is greater than both observed proportions.
  Such effects may easily lead to nonsensical inferences; for example, a meta-analysis of four proportions ($\frac{a}{N}=\frac{1}{10}$, $\frac{10}{100}$, $\frac{100}{1000}$, $\frac{1000}{10\,000}$) may yield a confidence interval excluding the common proportion of~$10\%$ (when using the harmonic mean sample size for back-transformation; here: $N=36$).
  The above calculations are demonstrated in the supplemental \textsf{R}~code.

  As previously pointed out by Schwarzer \emph{et~al.} (2019),\citep{SchwarzerEtAl2019} another consequence is that for certain (small or large) $\theta$~values no corresponding proportion~$p$ exists. 
  For example, in Figure~\ref{fig:example} a value of $\theta=0.015$ cannot be mapped to a corresponding proportion~$p$ for $N=676$ (blue line); sometimes such cases are then pragmatically mapped to $p=0$.
  

\section{Discussion and conclusions}
  All the above issues do not pose a problem in the context in which the transform was originally proposed, namely, the investigation of a \emph{single} binomial trial,\citep{FreemanTukey1950} but it means that proportions from differing sample sizes simply are not transformed to a common scale.
  While the double arcsine transformation may have good variance-stabilizing properties, it clearly has also some properties that make it unsuitable for meta-analysis.
  Use of the transform in a meta-analysis does not only lead to occasional erratic behaviour, but also to generally questionable inference.
  Note that the problems arise once the (forward) transformation is done, 
  and it is not a matter of improving upon the inverse transformation (or finding a suitable $N$~value), as has previously been suggested;\citep{DoiXu2021}  any back-transformation (seemingly) avoiding pathologies may in fact only be obscuring the problems.
  While one might argue that such issues may be less important when all sample sizes are similar, or when all sample sizes are large, it is also important to note that these issues do not arise for the obvious alternative of the closely related plain (``single'') arcsine transform.

  A number of more appropriate alternative approaches of differing technical complexity are available, as has been discussed e.g. by Schwarzer \emph{et~al.} (2019)\citep{SchwarzerEtAl2019} and Lin and Chu (2020).\citep{LinChu2020}
  A very obvious one among these may be the closely related, and in many ways simpler (``single'') arcsine transform, which also constitutes the limiting case of the double arcsine transformation for large sample sizes.  Recent empirical and numerical comparisons of meta-analysis methods for proportions did not indicate any advantages of the double arcsine transform over the plain (``single'') arcsine transform.\citep{LinXuChu2022,LinChu2020}
  If a two-stage approach approach, i.e, the combination of proportion estimates on a transformed scale, is desired, then alternative transformations should be considered, an obvious candidate being the (``single'') arcsine transform. In case there are still concerns regarding the quality of the approximations involved, one-stage models avoiding the intermediate transformation step should be used.\citep{SchwarzerEtAl2019,LinChu2020}
  
  In summary, while the double arcsine transform is well-suited for the originally intended purpose of providing a variance stabilizing transformation for a (single) proportion (and in this regard is superior to the arcsine transformation), it is clearly not appropriate for meta-analysis of \emph{several} binomial probability estimates.

\section*{Acknowledgment}
Support from the \emph{Deutsche Forschungsgemeinschaft (DFG)} is gratefully acknowledged (grant number \mbox{FR~3070/3-1}).

\section*{Conflicts of interest}
  The authors have declared no conflict of interest.

\section*{Data availability}
  The data that supports the findings of this study are available in the supplemental material of this article.

\section*{Highlights}
\paragraph{What is already known:}
\begin{itemize}
  \item Meta-analyses commonly involve transformations for endpoints or effect measures.
  \item Use of the variance-stabilizing double-arcsine transforms has been proposed in the context of meta-analyses of proportions.
  \item Some erratic behaviour of the double-arcsine transform has been observed previously.
\end{itemize}

\paragraph{What is new:}
\begin{itemize}
  \item We demonstrate why a double-arcsine transformation is not suitable for meta-analysis.
\end{itemize}

\paragraph{Potential impact for readers outside the authors’ field:}
\begin{itemize}
  \item While meta-analyses of proportions are common in many application areas, transformations of effect measures should in general be carefully selected to avoid paradoxical behaviour, in particular the double-arcsine transformation should not be used. Alternatives are available including the arcsine transformation or one-stage approaches.
\end{itemize}

\bibliography{literature}

\end{document}